\newcommand{\be}{\begin{equation}}
\newcommand{\bea}{\begin{eqnarray}}
\newcommand{\ee}{\end{equation}}
\newcommand{\eea}{\end{eqnarray}}
\newcommand{\nn}{\nonumber}

\newcommand{\qa}{\alpha}
\newcommand{\qb}{\beta}
\newcommand{\qg}{\gamma}

\newcommand{\qd}{\delta}

\newcommand{\qe}{\varepsilon}

\newcommand{\qy}{\theta}

\newcommand{\qk}{\kappa}
\newcommand{\ql}{\lambda}

\newcommand{\qs}{\sigma}

\newcommand{\qf}{\varphi}
\newcommand{\qF}{\Phi}

\newcommand{\qj}{\psi}
\newcommand{\qJ}{\Psi}

\newcommand{\rd}{{\rm d}}

\newcommand{\fr}[2]{{\textstyle \frac{#1}{#2}}}

\newcommand{\EE}{{\mathbb E}}

\newcommand{\bits}{ \{0,1\} }
\newcommand{\Hmin}{{\sf H}_{\rm min}}
\newcommand{\cC}{{\mathcal C}}

\newcommand{\cH}{{\mathcal H}}
\newcommand{\cI}{{\mathcal I}}

\newcommand{\cL}{{\mathcal L}}
\newcommand{\cM}{{\mathcal M}}

\newcommand{\cO}{{\mathcal O}}

\newcommand{\cR}{{\mathcal R}}
\newcommand{\cS}{{\mathcal S}}

\newcommand{\pr}{{\rm Pr}}

\newcommand{\isdef}{\stackrel{\rm def}{=}}

\newcommand{\pmax}{p_{\rm max}}

\newcommand{\ket}[1]{| #1 \rangle}
\newcommand{\bra}[1]{\langle #1 |}
\newcommand{\bigket}[1]{\Big| #1 \Big\rangle}
\newcommand{\inprod}[2]{\langle #1 | #2 \rangle}

\newcommand{\rej}{{\tt REJ}}

\newcommand{\zeroacc}{\mbox{0-{\tt ACC}}}
\newcommand{\oneacc}{\mbox{1-{\tt ACC}}}

\documentclass[10pt]{article}

\newtheorem{theorem}{Theorem}[section]

\newtheorem{lemma}[theorem]{Lemma}
\newtheorem{definition}[theorem]{Definition}
\newtheorem{corollary}[theorem]{Corollary}
\newtheorem{proposition}[theorem]{Proposition}
\newtheorem{conjecture}[theorem]{Conjecture}

\usepackage{graphics}
\usepackage{graphicx}
\usepackage{url}
\usepackage{amsmath}
\usepackage{amssymb}
\usepackage{bbold}
\usepackage{a4wide}
\usepackage{enumitem}

\begin{document}

\setlength{\parindent}{0mm}

\title{Quantum digital signatures with smaller public keys}

\author{
Boris \v{S}kori\'{c}
}

\date{ }

\maketitle

\begin{abstract}
\noindent 
We introduce a variant of quantum signatures in which nonbinary symbols are signed instead of bits.
The public keys are fingerprinting states, just as in the scheme of Gottesman and Chuang \cite{GC2001},
but we allow for multiple ways to reveal the private key partially.
The effect of this modification is a reduction of the number of qubits expended per message bit.
Asymptotically the expenditure becomes as low as one qubit per message bit.
We give a security proof, and we present numerical results
that show how the improvement in public key size depends on the message length.
\end{abstract}

%=============================  Intro ===========================================
\section{Introduction}

\subsection{Quantum signatures; unconditional security}

Digital signatures and Public Key Infrastructure (PKI) form the cornerstone
of our `open' digital world;
they allow people to verify the origin and integrity of data received from
new communication partners, in an almost entirely non-interactive (`offline') way and based merely 
on a small number of public keys stored locally.

In a typical signature scheme each user owns a private key $s$, which is kept secret,
and the related public key $p$, which is published. 
The public key is easily computed from the private key,
but the reverse computation is difficult because it involves a hard problem such as
factorisation, discrete logarithms, learning with errors, or a shortest vector problem.
Signing is an operation that takes as input $s$ and a message $m$, and outputs
a signature~$z$. 
Verification has the triplet $(m,p,z)$ as input, and produces a yes/no output,
where `yes' indicates that the signature $z$ is consistent with $m$ and~$p$.
A signature scheme has to satisfy three security properties:
(i) Unforgeability.
For someone who does not hold $s$ it is prohibitively difficult to create such a valid triplet;
(ii) Non-repudiation.
If a valid triplet $(m,p,z)$ is observed,
then the party associated with $p$ cannot deny that it has created the triplet and hence
endorses the message~$m$;
(iii) Transferability. 
If a verifier accepts a signature, he is confident that any other verifier will also accept it.

The main weakness of digital signature schemes is their
reliance on a difficult computational problem, whose hardness is impossible to prove.
For this reason alternative schemes have been studied 
\cite{CR1990,HSZI2000,SS2011}
that offer {\em unconditional security}.
These works have a number of disadvantages in common.
They have to work with a fixed set of participants, 
and they involve a large amount of communication.
Furthermore, they require either a trusted third party or secret channels between pairs of participants.

Gottesman and Chuang \cite{GC2001} introduced {\em quantum digital signatures}, 
which are unconditionally secure and alleviate some of these disadvantages. 
The main idea is based on the observation that state preparation can be seen as
a one-way function. Consider a prover Peggy who gives a quantum state to a verifier Victor.
It is easy for Peggy to put a huge amount of information into a quantum state
but impossible for Victor to extract all of it.
It is also straightforward for Peggy to convince Victor that she knows exactly what the state is. 
From this unconditionally secure one-way function one can then build a Lamport-like 
\cite{Lamport1979} signature scheme.
In the Gottesman-Chuang scheme \cite{GC2001} 
(which we will abbreviate as `GC01')
the private key is the classical data that Peggy puts into quantum states;
the thus produced states are the public key.
Multiple instances of the public key are allowed to exist, and these are given to
the verifiers. 
It does not have to be fixed beforehand who the verifiers are,
and they do not have to communicate beforehand;
this flexibility is the main advantage of quantum signatures over the classical unconditionally-secure schemes.

In GC01
it is implicitly assumed that there exists some mechanism by which the verifiers 
can trust that the quantum states they receive ultimately originate from Peggy.
This mechanism must not rely on standard PKI with its computational assumptions but 
e.g.~on trusted point-to-point contacts.
The complications of such a key transport mechanism are a disadvantage compared to ordinary PKI.
A further disadvantage is of course the need for quantum memory at the verifiers' side, and for
quantum channels.

In 2014--2015 several versions of quantum signatures were introduced \cite{DWA2014,CRDWCAJB2014} that do not need quantum memory.
However, they have the disadvantage that all recipients of the public key\footnote{
Confusingly refered to as `signature'.
} need to participate
in the distribution stage of the protocol.

A review of quantum signatures was given in \cite{AA2015}.

%--------------------------------------------------------
\subsection{Our contribution}
\label{sec:results}

We introduce a new variant of Gottesman-Chuang like
quantum signatures (with quantum memory)
in which Peggy is able to `open' a public key in multiple ways, thus signing
a non-binary symbol instead of a bit.
Our public-key qudits are fingerprinting states \cite{BCWdW2001,GI2013}.
Our digital signature reveals only {\em a substring of the full string embedded in the public key};
the substring can be chosen in multiple different ways.
We show that this method reduces the amount of public-key material required for the signing of a message.
For the sake of efficiency
our scheme uses the idea suggested in \cite{GC2001} to work with codewords
instead of repeated public keys, but it does so with {\em non-binary} symbols.

The price to pay for revealing only partial information is that there is now a nonzero
error probability when verifying a legitimate qudit (compared to zero in \cite{GC2001}),
and furthermore forgery becomes slightly easier.
Nevertheless, the overall tradeoff between security and efficiency works in our favour:
at a given level of security (expressed as the gap between Peggy's and the adversary's success probability
to open a qudit)
our scheme spends fewer qubits per signed message bit than~\cite{GC2001}, approximately
$1+\frac{\log(T\log T)}{\log S}$, where $T$ is the number of verifiers 
and $S$ is the size of the alphabet
(see Section~\ref{sec:GCfinger}).
Asymptotically the size of the public key approaches as little as one qubit per signed message bit.
In contrast, GC01 needs at least $\approx \log(T\log T)$.

The outline of this paper is as follows.
In the preliminaries (Section~\ref{sec:prelim}) we introduce notation
and list a number of useful lemmas.
We briefly recapitulate the GC01 scheme \cite{GC2001} and
fingerprinting states \cite{BCWdW2001}.
In Section~\ref{sec:merit} we look at the relation between non-repudiation on the one hand
and correctness and security against forgery on the other hand. 
We discuss the difference between the true reject and false reject probability as
a performance indicator.
In Section~\ref{sec:analysisGC} we look at GC01 in more detail
and derive a lower bound on the number of qubits spent per signed message bit.
In Section~\ref{sec:ourscheme} we introduce our scheme,
and in Section~\ref{sec:analysis} we present the analysis.
We summarize in Section~\ref{sec:summary}.

%====================================================
\section{Preliminaries}
\label{sec:prelim}

%-------------------------------------------
\subsection{Notation, attacker model, and security definitions}
\label{sec:notation}

\underline{Notation}.
There are $T$ verifiers. 
We write $d$ for the dimension of the public-key Hilbert space.
We use the notation $[d]=\{0,\ldots,d-1\}$. 
A private key is a string $k\in\bits^d$.
Let $\cI\subset[d]$ be a subset.
We write $k_\cI$ for the substring $(k_i)_{i\in\cI}$ where ordering of $\cI$ is applied.
The complement of $\cI$ is denoted as $\cI^{\rm c}=[d]\setminus\cI$.

Our scheme signs non-binary symbols in an alphabet $\cS$ of size~$S$.
We write $\cS=\{0,\ldots,S-1\}$.

The Hamming weight of a binary string $x$ is denoted as $|x|$.
The bitwise XOR of binary strings $x$ and $y$ is written as $x\oplus y$.

The notation $h$ stands for the binary entropy function
$h(p)=p\log\fr1p+(1-p)\log\fr1{1-p}$.

\underline{Attacker model}.
The adversary has unlimited (quantum) computing power, as well as
measurement and state preparation equipment that is entirely without noise.
The adversary has no access to the labs of the other parties
(e.g. through side channels).

\underline{Security definitions}.

We work in the following setting.
Let $\cH$ be a Hilbert space.
Peggy has a private key $k$ which is a classical string.
She uses $k$ to create $T$ copies of a public key $\ket{P_k}\in\cH$.\footnote{
We assume that there is a mechanism for distributing public keys.
In this respect we do not deviate from the assumptions made in \cite{GC2001}.
}
A signature of a classical message $m\in\cM$  
is a classical string $r={\sf Sign}(k,m)$, $r\in\cR$ which is computed as a function of $k$ and~$m$.
Signature verification is an algorithm {\sf Verif} that acts on a state
$\ket P\in\cH$, a message $m\in\cM$ and a string $r\in\cR$, yielding outcome
$v={\sf Verif}(\ket P,m,r)\in\{\rej,\oneacc,\zeroacc \}$.
Here \rej\, stands for rejection;
\oneacc\, means that Victor considers the signature to be valid, and that he is confident that
any other verifier will also consider it to be valid;
\zeroacc\, means that Victor considers the signature to be valid, but is not sure about other verifiers.

\begin{definition}[Correctness]
\label{def:correctness}
We say that the signature scheme is {\em correct with error} $\qe$ if
\be
	\forall_{k,m}\;\;
	\pr\Big[{\sf Verif}\Big(\ket{P_k},m, {\sf Sign}(k,m)\Big)=\oneacc\Big] \geq 1-\qe.
\ee
\end{definition}

\begin{definition}[Security against forgery]
\label{def:forgery}

Let $k\in\bits^d$ be generated randomly,
and let $\ket{P_k}$ be the corresponding public key state.
Consider an adversary who has access to $\ket{P_k}^{\otimes T}$,
chooses one message $m\in\cM$ and receives the signature $r={\sf Sign}(k,m)$.
The adversary then outputs a pair $(m',r')$, with $m'\in\cM$, $r'\in\cR$.
We call the signature scheme $\qe$-secure against forgery if
\be
	\pr\Big[m'\neq m \;\wedge\; {\sf Verif}(\ket{P_k},m',r')\neq\rej\Big]\leq\qe.
\ee
Here the probability is taken over the random $k$, the adversary's random choices,
and the nondeterministic outcome of {\sf Verif}.
\end{definition}

\begin{definition}[Non-repudiation heuristic]
\label{def:nonrepudiation}
Let malicious Peggy pick any state $\ket\qJ\in\cH$, any message $m\in\cM$, and any string~$r\in\cR$;
these are given to the $T$ verifiers.
Let each verifier independently execute ${\sf Verif}(\ket\qJ,m,r)$.
Let $N_{\tt 1ACC},N_\rej$ denote the number of verifiers that get result
\oneacc, \rej\, respectively.
We call the signature scheme $\qe$-secure against repudiation if
\be
	\forall_{\qJ,m,r}\;\;
	\pr[N_{\tt 1ACC}\geq 1 \wedge N_\rej \geq 1] \leq \qe.
\label{repudprobdef}
\ee
\end{definition}
Def.\,\ref{def:nonrepudiation} does not allow malicious Peggy to
hand out different states to different verifiers, 
in contrast to
the repudiation attacker model in GC01 which allows more general (entangled) states that pass swap tests.
Hence Def.\,\ref{def:nonrepudiation} should be seen as a security heuristic and not
a full security definition.

%devifucation  (typo of verification)

%--------------------------------------------------------
\subsection{Tail bounds}

\begin{lemma}
\label{lemma:radius}
Let $r\leq\frac n2$. The following inequalities hold,
\be
	\frac{2^{nh(r/n)}}{\sqrt{8r(1-r/n)}}
	\leq\sum_{k=0}^r {n\choose k}\leq 2^{nh(r/n)}.
\ee
\end{lemma}
\underline{\it Proof:}
For the first inequality see e.g.~p.121 of~\cite{Ash}.
The second inequality is a special case of Chernoff-Hoeffding with probability parameter~$1/2$.
\hfill$\square$

\begin{lemma}[Chernoff bound]
\label{lemma:Chernoff}
Let $X=\sum_i X_i$ with $X_i\in\bits$ independent random variables. Let $\mu=\EE X$.
Then for any $\qd>0$ it holds that
\be
	\pr[X \geq \mu+\mu\qd] \leq e^{-\fr1{2+\qd} \qd^2\mu}.
\ee
\be
	\pr[X \leq \mu-\mu\qd] \leq e^{-\fr12 \qd^2\mu}.
\ee
\end{lemma}

%--------------------------------------------------------
\subsection{The Gottesman-Chuang scheme \cite{GC2001}}
\label{sec:prelimGC}

We briefly summarize the efficient version of GC01,
using codewords, as presented in Section~8 of their paper.

The message to be signed is $x\in\bits^K$.
It is encoded into a codeword $c\in\bits^N$.
The distance of the code is~$M$.
The private key is $k=(k^0_j,k^1_j)_{j=1}^N$, with $k_j^{0/1}\in\bits^L$.
The public key consists of $2N$ $d$-dimensional qudit states 
$(\ket{P^0_j})_{j=1}^N$, $(\ket{P^1_j})_{j=1}^N$,
with $\ket{P^b_j}= \ket{F(k^b_j)}$,
where $F$ denotes some method of embedding the string $k^b_j$ into the qudit.
There are $T$ copies of the public key.
The parameter $\qd$, which depends on the embedding method and the dimension of the Hilbert space, is defined as 
\be
	\qd=\max_{k,k': k\neq k'} \Big| \inprod{F(k')}{F(k)} \Big|.
\label{defdelta}
\ee

Peggy's signature of the string $x$ consists of the 
private keys $(k^{c_j}_j)_{j=1}^N$.
Signature verification is done by projecting the state
$\ket{P^{c_j}_j}$ in possession of the verifier onto the direction $\ket{F(k^{c_j}_j)}$, 
for each $j\in\{1,\cdots,N\}$, and counting the number of `0' results (`$z$').
There are two threshold parameters, $z_{\rm acc},z_{\rm rej}$, with $z_{\rm acc}<z_{\rm rej}\leq M/2$.
If $z\leq z_{\rm acc}$ then the result of the verification is {\tt 1-ACC};
if $z\geq z_{\rm rej}$ then it is {\tt REJ}. 
In between, the result is {\tt 0-ACC}.

Regarding forgery the following result was shown.
An adversary who holds all $T$ copies of the public key can learn no more than
$T\log d$ bits of information about the private key $k^{c_j}_j\in\bits^L$.
The forgery probability for a single qudit is therefore upper bounded by the following value,
\be
	p_{\rm forge1}^{\rm GC}=\frac1{2^{L-T\log d}}+(1-\frac1{2^{L-T\log d}})\qd^2.
\label{pforge1GC}
\ee

% Swap test

%--------------------------------------------------------
\subsection{Fingerprinting states}

Quantum fingerprinting was introduced in 
\cite{BCWdW2001}
as a way to do string equality testing based on 
a representation that is exponentially smaller than the classical string.
Let $x\in\bits^d$, $x=(x_j)_{j=0}^{d-1}$.
Let $\ket0,\ldots,\ket{d-1}$ be an orthonormal basis of a $d$-dimensional Hilbert space~$\cH$.
The {\em fingerprinting state} $\ket{\mu(x)}$ for the string $x$ is the following state in $\cH$,  
\be
	\ket{\mu(x)}\isdef \frac1{\sqrt d}\sum_{j=0}^{d-1}(-1)^{x_j}\ket j.
\label{deffingerprint}
\ee
These states have been used for various other purposes, e.g. noise-tolerant QKD \cite{SYK2014}.
It was proposed by Gottesman and Chuang to use fingerprinting states as the embedding mechanism `$F$'
in their quantum signature scheme.

\begin{lemma}
\label{lemma:inprodmu}
Let $x,y\in\bits^d$.
The inner product of the two fingerprint states $\ket{\mu(x)}$ and $\ket{\mu(y)}$
is given by
\be
	\inprod{\mu(y)}{\mu(x)} = 1-2\frac{|x\oplus y|}{d}.
\ee
\end{lemma}
\underline{\it Proof:}
From the definition (\ref{deffingerprint}) we get
$\inprod{\mu(y)}{\mu(x)} =\fr1d\sum_{j,\ell=0}^{d-1}(-1)^{y_\ell+x_j}\inprod{\ell}{j}$
$=\fr1d\sum_{j=0}^{d-1}(-1)^{y_j+x_j}$
$=\fr1d\sum_{j=0}^{d-1}(1-2y_j\oplus x_j)$
$=\fr1d(d-2|y\oplus x|)$.
\hfill$\square$

%========================================================================
\section{Figures of merit}
\label{sec:merit}

Given a state $\ket P\in\cH$, a message $m\in\cM$ and an alleged signature $r\in\cR$,
let $Q_{\rm R},Q_0,Q_1$ denote the probability that ${\sf Verif}(\ket P,m,r)$
yields outcome \rej, \zeroacc, \oneacc\, respectively.
Let $N_\rej,N_{\tt 0ACC},N_{\tt 1ACC}$ denote the number of verifiers 
who get those outcomes.

\begin{lemma}
\label{lemma:repud}
The repudiation probability (\ref{repudprobdef}) can be expressed as
\be
	P_{\rm repud}\isdef
	\pr[N_{\tt 1ACC}\geq 1 \wedge N_\rej \geq 1] = 1-(1-Q_{\rm R})^T - (1-Q_1)^T +Q_0^T.
\label{repudlemma}
\ee
\end{lemma}

\underline{\it Proof:}
The outcome for each of the $T$ verifiers is independent and follows the same distribution
$(Q_{\rm R},Q_0,Q_1)$. 
The left hand side of (\ref{repudlemma}) can be written as a partial sum over the multinomial probability distribution,
$\sum_{a=1}^{T-1}\sum_{b=1}^{T-a}\frac{T!}{a! b! (T-a-b)!}Q_{\rm R}^a Q_1^b Q_0^{T-a-b}$,
which can be rewritten as
$\sum_{a=1}^{T-1}{T\choose a}Q_{\rm R}^a\sum_{b=1}^{T-a} {T-a\choose b}  Q_1^b Q_0^{T-a-b}$
$=\sum_{a=1}^{T-1}{T\choose a}Q_{\rm R}^a[(Q_0+Q_1)^{T-a}-Q_0^{T-a}]$.
Finally we use the binomial sum rule twice, subtracting the $a=0$ and $a=T$ terms. 
\hfill$\square$

\begin{corollary}
\label{corol:repud}
The following inequalities hold
\bea
	P_{\rm repud} &\leq& 1-\Big(1-\min(Q_{\rm R},Q_1)\Big)^T
	\leq 1-\Big(\max(Q_{\rm R},Q_1)\Big)^T
\label{Prepudineq1}
	\\
	P_{\rm repud} &\leq& T\cdot \min(Q_{\rm R},Q_1).
\label{Prepudineq2}
\eea
\end{corollary}
\underline{\it Proof:}
The first inequality in (\ref{Prepudineq1}) is obtained from (\ref{repudlemma}) by 
using $Q_0^T \leq (1-Q_{\rm R})^T$ and $Q_0^T \leq (1-Q_1)^T$.
The second one follows from $Q_1+Q_{\rm R}\leq 1$.
The inequality (\ref{Prepudineq2}) follows from the first expression in (\ref{Prepudineq1})
by using $(1-x)^T \geq 1-Tx$.
\hfill$\square$

\begin{lemma}
\label{lemma:impliesrepud}
Consider a prover who hands out a state in $\cH$ which is indeed the public key
$\ket{P_k}$ for some private key $k$.
If a signature scheme is correct with $\qe_1$-error and is $\qe_2$-secure against forgery,
then for any $m\in\cM$, $r\in\cR$ the repudiation probability $P_{\rm repud}$ is upper bounded by 
$T\cdot\max(\qe_1,\qe_2)$.
\end{lemma}

\underline{\it Proof:}
We distinguish between two cases, 
(i) $r$ is a correct signature, and (ii) $r$ is not a correct signature.
For the first case we use (\ref{Prepudineq2}) to write
$P_{\rm repud}\leq TQ_{\rm R}\leq T(1-Q_1)$, with $Q_1\geq 1-\qe_1$.
Similarly, in the second case we use 
(\ref{Prepudineq2}) to write
$P_{\rm repud}\leq TQ_1\leq T(1-Q_{\rm R})$, with $Q_{\rm R}\geq 1-\qe_2$.
\hfill$\square$

\vskip2mm

Note that Lemma~\ref{lemma:impliesrepud} does not imply non-repudiation as
defined in Def.~\ref{def:nonrepudiation}.
The reason is that correctness and security against forgery are defined only
for quantum states $\ket P$ which are a proper public key $\ket{P_k}$ for some private key $k$,
whereas the definition of non-repudiation allows Peggy to distribute {\em any} state in~$\cH$.

On the other hand, Lemma~\ref{lemma:impliesrepud} provides a guideline on
how the correctness error and the security against forgery should be tuned if
one aims at a certain level of non-repudiation.

We add a superscript `genuine' or `forgery' on the probabilities $Q_{\rm R},Q_0,Q_1$
to distinguish between the two cases.

In the schemes that we focus on in this paper, we have $\cH=\cH_1^{\otimes N}$.
The $\cH_1$ is referred to as a qudit space.
The verifier performs a binary projective measurement on each of the $N$ individual qudits,
e.g.\,the projection onto $\ket{F(k_j^{c_j})}$ in GC01 (Section \ref{sec:prelimGC}),
and gets a tally $z\in\{0,\ldots,N\}$ of how many errors occur.
(By `error' we mean that a qudit does not pass verification, i.e. the projection yields `0'.)
Let $G$ denote the per-qudit error probability in case of a {\em genuine} signature, and $J$ in case of a mismatch
between the signature and the quantum state in~$\cH_1$.
The relevant quantity for the correctness property is the error tally in case of a genuine signature,
\be
	Q_1^{\rm genuine}=\pr[Z\leq z_{\rm acc}|{\rm genuine}] = \sum_{z=0}^{z_{\rm acc}}{N\choose z}G^z (1-G)^{N-z}.
\label{Q1G}
\ee
If a lower bound $Q_1^{\rm genuine}\geq1-\qe_c$ can be proven, then the scheme is
`correct with error $\qe_c$' as specified in Def.\,\ref{def:correctness}.
For the security against forgery the relevant quantity is
\be
	Q_{\rm R}^{\rm forgery} = \pr[Z\geq z_{\rm rej}|{\rm forgery}].
\ee
If a lower bound $Q_{\rm R}^{\rm forgery}\geq 1-\qe_f$ can be proven,
then the scheme is `$\qe_f$-secure against forgery' as specified in Def.\,\ref{def:forgery}.
Unfortunately, a relation such as (\ref{Q1G}) does not necessarily exist
between $Q_{\rm R}^{\rm forgery}$ and $J$, as we will see in Section~\ref{sec:analysis},
because in case of a forgery not all positions $1,\ldots,N$ have to be a mismatch.
However, it is clear that $J>G$ is a necessary condition in order to have a working signature scheme.
Let $N'$ ($N'<N$) be the number of positions where a forgery causes a mismatch in the quantum state.
The error tally $z^{\rm forgery}$ is peaked around $N' J+(N-N')G$, whereas $z^{\rm genuine}$ is peaked around
$NG$.
In order to distinguish between genuine signatures and forgeries, the scheme must have a significant
distance between the thresholds $z_{\rm rej}$ and $z_{\rm acc}$, which implies the condition
$N' J+(N-N')G > NG \Leftrightarrow N'(J-G)>0$. 
The distance grows with $N'$.
A large distance is necessary in order to reduce the overlap between the right tail of $z^{\rm genuine}$
and the left tail of $z^{\rm forgery}$.

Because of the structure discussed above, we adopt the `{\bf gap}' $J-G$ as one of the central figures of merit.
The other figures of merit are the length of the code ($N$) and the total size of the public key
expressed in qubits ($N\log d$).

\begin{table}
\centering
\caption{\it Notation.}
{\small
\begin{tabular}{|l|l|}
\hline
$\qa$ & $=(d-\ell)/d$. Relative size of non-revealed substring. \\
\hline
$\qb$ & In GC01: bit error rate that can be corrected by the code \\
\hline
$c$ & codeword. $c\in\cS^N$ \\
\hline
$d$ & Dimension of the Hilbert space. String length. \\
\hline
$\qd$ & In GC01: $|\inprod{P(k)}{P(k')}|\leq\qd$ \\
\hline
$\qf$ & Small parameter. $\qf=d^{-1}T\log d$ \\
\hline
$G$ & single-qudit false reject probability \\
\hline
$\cI$ & Set containing the indices of non-revealed positions. $|\cI|=d-\ell$.  \\
\hline
$J$ & single-qudit true reject probability \\
\hline
$k$ & Private key. In our scheme $k\in\bits^d$.  \\
\hline
$\qk$ & Substring of private key. $\qk\in\bits^\ell$\\
\hline
$K$ & message length\\
\hline
$L$ & Private key size in GC01.\\
\hline
$\ell$ & number of revealed positions \\
\hline
$\ket{\mu(z)}$ & Fingerprinting state for string $z\in\bits^d$. \\
\hline
$n$ & size of public key in qubits. $n=\log_2 d$ \\
\hline
$N$ & codeword length \\
\hline
$\ket P$ & Public key.  
\\
\hline
$\ket\qj$ & state derived from index set $\cI$ and substring $\qk$ \\
\hline
$Q_{\rm R},Q_0,Q_1$ & probability of Reject, ACC-0, ACC-1 \\
\hline
$s$ & Symbol to be signed. $s\in\cS$. \\
\hline
$S$ & Alphabet size. $S=1/\qa$. \\
\hline
$\cS$ & Alphabet. $|\cS|=S$. $\cS=\{0,\ldots,S-1\}$. \\
\hline
$T$ & Number of verifiers. (Number of copies of each public key) \\
\hline
$\qy$ & bit error rate that can be corrected \\
\hline
$x$ & Message. $x\in\cS^K$. \\
\hline
\end{tabular}
}
\end{table}

%\clearpage

%========================================================================
\section{Analysis of the Gottesman-Chuang scheme}
\label{sec:analysisGC}

%-----------------------------------------------------
\subsection{Gottesman-Chuang with fingerprinting states}
\label{sec:GCfinger}

We present an analysis of the GC01 scheme that explicitly writes out 
a number of parameters that were not worked out in detail in~\cite{GC2001}.
We consider the efficient implementation with codewords and fingerprinting states.
We look only at long-message asymptotics.

The quantities of interest are
(i)
The number of qubits spent on signing a whole message;
(ii)~the value of the parameter $\qd$;
(iii)
the ensuing single-bit forgery success probability, and
(iv) the minimum codeword length (number of public keys) required to upper-bound the overall
message forgery probability to some fixed value.

In particular, the number of spent qubits and the single-bit forgery probability are important 
as a benchmark for evaluating the performance of our own scheme. 

\vfill

\newpage

\underline{The number of spent qubits per message bit}.\newline
The error-correcting code is a binary code (`$\cC_1$') with message size $K$, codeword size $N$, and distance~$M$.
Then $\qb\isdef\frac M{2N}$ is the error rate that can be corrected.
Asymptotically for large messages it holds\footnote{
Ref.\cite{BKB2004} gives the following result for the length of the syndrome,
$nh(\qb)+\sqrt{n}\qF^{\rm inv}(10^{-6})\sqrt{\qb(1-\qb)}\log\frac{1-\qb}{\qb}$, 
where $\qF$ is defined as $\qF(z)\isdef \int_z^\infty(2\pi)^{-1/2}\exp[-x^2/2]\rd x$.
}
 that $\frac KN\approx 1-h(\qb)$.

The number of public keys required for signing the codeword is~$2N$.
Each public key state comprises $\log d$ qubits. 
Hence the number of qubits involved in signing a $K$-bit message is $2N\log d$.
\be
	\mbox{GC01 \#qubits per message bit}= \frac{2N\log d}K \approx \frac{2\log d}{1-h(\qb)}. 
\label{GCqubits}
\ee
Note that only half of the public keys are `opened'.
The expenditure of qubits may be halved if there is a way of re-using the unused public keys.

\vskip2mm
\underline{The parameter $\qd$}.\\
A code $\cC_2$ is used in the embedding, with message length $L$ and codeword length~$d$.
The key $k\in\bits^L$ ($L<d$) is embedded in the public key as as the fingerprinting state
of a $d$-bit codeword in $\cC_2$.
We denote the correctable error rate of this code as $\qg$, with (again asymptotically)
$\frac Ld\approx 1-h(\qg)$.
In order to compute the parameter $\qd$ as defined in (\ref{defdelta})
we consider two keys $k,k'$ which differ only by a single bit flip.
Their codewords differ in $2\qg d$ bits.
From Lemma~\ref{lemma:inprodmu} it then follows that
\be
\label{deltagamma}
	\qd=1-4\qg.
\ee

\vskip2mm

\underline{The single-bit forgery probability}.\\
Substitution of (\ref{deltagamma}) into (\ref{pforge1GC}) yields
\bea
	p_{\rm forge1}^{\rm GC} &=& 
	(\frac12)^{d[1-h(\qg)]-T\log d} +\Big\{1-(\frac12)^{d[1-h(\qg)]-T\log d}\Big\}(1-4\qg)^2
	\nn\\ &=&
	1 -8\qg(1-2\qg)\Big\{1-(\frac12)^{d[1-h(\qg)]-T\log d}\Big\}
\eea
where we have expressed $L$ in terms of $d$ and~$\qg$.

The error probability $J$ introduced in Section~\ref{sec:merit} equals $1-p_{\rm forge1}$.
Furthermore, in GC01 the error probability $G$ vanishes, $G^{\rm GC}=0$.
Hence the `gap' figure of merit is given by
\be
	J^{\rm GC} = 8\qg(1-2\qg)\Big\{1-(\frac12)^{d[1-h(\qg)]-T\log d}\Big\}.
\ee

Note that the condition $L>T\log d$ has to hold, which translates to
$\frac d{\log d}>\frac T{1-h(\qg)}$. 
Hence there is a lower bound $d_{\rm min}^{\rm GC}$ on the dimension of the Hilbert space,
dictated mostly by the parameter $T$. 
For small $\qg$ this bound is
\be
	d_{\rm min}^{\rm GC} \approx T\log T.
\label{dminGC}
\ee
Substitution of (\ref{dminGC}) into (\ref{GCqubits}) yields the following 
approximation for the minimum size of the public key,
\be
	\mbox{Small }\qg: \quad 
	\mbox{GC01 \#qubits per message bit} \gtrapprox  \frac{2\log (T\log T)}{1-h(\qb)}. 
\label{GCsmallgamma}
\ee

\vskip2mm

\underline{Minimum codeword length}.\\
Let $Q_{\rm R}^{\rm forgery}$ be a target value that we want to achieve
regarding the forgery detection probability for a whole message.
The easiest forgery is to flip a single bit in the message~$x$.
This causes $2\qb N$ flips in the codeword~$c$.
The forger has probability of at most $p_{\rm forge1}^{\rm GC}$
to repair a flip. 
Hence the expected number of leftover `wrong' bits counted by the tally $z$
is $\EE z=2\qb N(1-p_{\rm forge1}^{\rm GC})=2\qb N J^{\rm GC}$.
The threshold $z_{\rm rej}$ has to be set as
\be
	z_{\rm rej}=2\qb N J^{\rm GC}-2\sqrt{\qb N J^{\rm GC}}\sqrt{\ln (1-Q_{\rm R}^{\rm forgery})^{-1}}
\ee
(or smaller).
From Lemma~\ref{lemma:Chernoff} it follows that this setting indeed yields the
correct bound on the forgery probability.
The requirement $z_{\rm rej}>0$ leads to the condition $N\geq N_{\rm min}^{\rm GC}$, with
\be
	N_{\rm min}^{\rm GC}=\frac{\ln (1-Q_{\rm R}^{\rm forgery})^{-1} }{\qb J^{\rm GC}}.
\ee
Choosing a small $\qg$ on the one hand reduces the dimension $d$,
but on the other hand increases~$N_{\rm min}^{\rm GC}$. 
Similarly, setting $\qb$ small reduces the number of qubits spent per message bit (\ref{GCqubits})
but increases~$N_{\rm min}^{\rm GC}$.
For $\qg\ll 1$ we have $N_{\rm min}^{\rm GC}\propto \frac1{\qb\qg}$.

%----------------------------------------------------------------
\subsection{Gottesman-Chuang with low-dimensional embedding}
\label{sec:GClowdim}

In \cite{GC2001} the possibility was mentioned of embedding $k\in\bits^L$
into a single qubit ($d=2$).
Though possible, it has the drawback that the forgery error probability $J$ gets 
exponentially close to $1$, namely
$J=\cO(2^{-L})$.
The $L$ is lower bounded as $L > T\log d=T$, which yields
$J=\cO(2^{-T})$.

We briefly comment on the possibility of embedding $k$ into a Hilbert space of dimension $d$
larger than~2 but much smaller than (\ref{dminGC}).
Optimal spreading of states is equivalent to distributing $2^L$ points equally over a hypersphere of
dimension $\qs=2d-2$.
Each point dominates a $\qs$-dimensional solid angle of order $2^{-L}$,
and hence the angle between neighbouring points is $\cO(2^{-L/\qs})$.
The parameter $\qd$ is the cosine of this angle.
Substitution into (\ref{pforge1GC}) gives
$1-p_{\rm forge1}=\cO(2^{-2L/\qs})=\cO(2^{-L/[d-1]})$
$<\cO(2^{-\frac{T\log d}{d-1}})$.
At fixed small $d$ the distinction between Peggy and the attacker is exponentially small
in~$T$.  

Because of the exponentially small $J$ value in low-dimensional embedding,
we will use the fingerprinting-based version of GC01
as our benchmark.

%========================================================================
\section{Our protocol for signing a nonbinary string}
\label{sec:ourscheme}

\subsection{Intuition}

We propose a scheme that is similar to the Gottesman-Chuang scheme with codewords and fingerprinting states, 
but which allows Peggy to `open'
a public key in $S$ different ways. 
The choice how to open corresponds to signing a symbol $s\in\cS=\{0,\ldots,S-1\}$.
Peggy's private keys are $(k^i)_{i=1}^N$ with $k^i\in\bits^d$. 
A public key $\ket{P_i}$ consists of the fingerprinting state $\ket{\mu(k^i)}$,
i.e.~without the use of an error-correcting code `$\cC_2$' in the embedding.

Peggy `opens' the public key $\ket{P_i}$
by revealing a length-$\ell$ substring of~$k^i$.
The choice of non-revealed positions encodes the symbol that is to be signed.
The verification step is to project $\ket{P_i}$ onto the average 
of all the fingerprinting states consistent with the revealed substring.

The intuition is that, on the one hand, $\ell$ is large enough such that by revealing $\ell$ bits of $k^i$
Peggy really proves that she knows $k^i$, while on the other hand
the number of non-revealed bits
($d-\ell$) is large enough to prevent forgeries.

%------------------------------------------------------------
\subsection{Substring positions}
\label{sec:I}

We set $S,\ell,d$ such that $S(d-\ell)=d$.
For $s\in\{0,\ldots,S-1\}$ we define disjoint subsets $\cI(s)\subset[d]$ 
with $|\cI(s)|=d-\ell$,
\be
	\cI(s) \isdef \{ s(d-\ell),\ldots,(s+1)(d-\ell)-1 \}.
\ee
The subset $\cI(s)$ points at the non-revealed positions in the private key. 
For convenience we define a `small' parameter $\qa$ as $\qa\isdef 1/S$.

%------------------------------------------------------------
\subsection{Protocol steps}
\label{sec:steps}

\underline{System setup}\\
Choose message length $K$, alphabet size $S$ and Hilbert space dimension $d$,
with $S$ dividing $d$.
The parameters $\ell$ and $\qa$ follow as $\ell=d-d/S$, $\qa=1/S$.
Choose an error correcting code $\cC$ (over the alphabet $\cS$) 
with codeword size $N$ that can correct symbol error rate~$\qy$.
Set error tally thresholds $z_{\rm acc},z_{\rm rej}\in\{0,\ldots,N\}$, with
$\qa N\leq z_{\rm acc}<z_{\rm rej}\leq 2\qy N$.

\underline{Protocol}
\begin{enumerate}[leftmargin=5mm,itemsep=0mm]
\item
{\bf Distribution of public keys}.
Peggy generates private keys $(k^i)_{i=1}^N$,
$k^i\in\bits^d$. 
For $i\in\{1,\ldots,N\}$
she prepares $T$ copies of the public key $\ket{P_i}\isdef\ket{\mu(k^i)}$.
Each verifier receives $\ket{P_1},\ldots,\ket{P_N}$.
\item
{\bf Signing}.
Peggy announces a message $x\in\cS^K$. 
She encodes $x$ to a codeword~$c\in\cS^N$ in the code~$\cC$.
She signs each individual symbol of~$c$ as follows. 
To sign $c_i\in\cS$ she announces the substring $\qk_i\isdef k^i_{[d]\setminus\cI(c_i)}$,
i.e.\,$k^i$ without the positions $\cI(c_i)$.
\item
{\bf Verification}.
Victor receives possibly corrupted data $x'\in\cS^K$ and $(\qk_i')_{i=1}^N$, $\qk_i'\in\bits^\ell$.
He performs the following actions.
Encode $x'$ to $c'\in\cS^N$.
For all $i\in[N]$ compute the normalized vector $\qj_i$ as
\be
	\ket{\qj_i}\propto \sum_{k\in\bits^d:\; k_{[d]\setminus\cI(c_i')}=\qk_i'}\bigket{\mu(k)}.
\label{verifpsi}
\ee 
For all $i\in[N]$
apply the projective measurement $\ket{\qj_i}\bra{\qj_i}$ on $\ket{P_i}$.
Let $z\in\{0,\ldots,N\}$ be the tally of `0' outcomes. 
If $z\leq z_{\rm acc}$ then the result of the verification is {\tt 1-ACC};
if $z\geq z_{\rm rej}$ then the result is {\tt REJ}. 
In between, the result is {\tt 0-ACC}.
\end{enumerate}

%==============================================================
\section{Analysis of the proposed scheme}
\label{sec:analysis}

%------------------------------------------------------------
\subsection{False reject probability per symbol ($G$)}

We look at the case where a public key $\ket{P_i}$ is unchanged after the Verifier receives it, 
and Peggy correctly signs. 
We compute the probability that the verification of one symbol fails.

\begin{lemma}
\label{lemma:truepos}
When the projection onto $\ket\qj\bra\qj$ is done in step~3,
the probability of a `1' outcome in a qudit is given by
\be
	\pr\Big[{\rm projection\; onto}\;\ket\qj\bra\qj\Big] = \frac\ell d=1-\qa.
\ee
\end{lemma}

\underline{\it Proof:}
Without loss of generality we take $c_i=S-1$. 
Then $\cI(c_i)$ equals $\{\ell,\ldots,d-1\}$;
the state $\qj$ is computed as
\bea
	\ket\qj &\propto&
	\sum_{a\in\bits^{d-\ell}}\bigket{\mu(\qk||a)}
	\nn\\	&=&
	\sum_{a\in\bits^{d-\ell}} \frac1{\sqrt d}\left[
	\sum_{j=0}^{\ell-1}(-1)^{\qk_j}\ket j
	+\sum_{j=\ell}^{d-1}(-1)^{a_{j-\ell}}\ket j
	\right]
	\nn\\ &=&
	\frac{2^{d-\ell}}{\sqrt d} \sum_{j=0}^{\ell-1}(-1)^{\qk_j}\ket j.
\eea
For general $\cI$ and $\qk$ we have
\be
	\ket{\qj(\cI,\qk)} = \frac1{\sqrt\ell} \sum_{j=0}^{\ell-1}(-1)^{\qk_j}\bigket{([d]\setminus\cI)_j}.
\ee
The probability of outcome `1' is computed as the square of the following inner product,
\be
	\inprod{\mu(k)}{\qj(\cI,k_\cI)} 
	=
	\frac1{\sqrt{d\ell}}\sum_{j'=0}^{d-1}\sum_{j=0}^{\ell-1}(-1)^{k_{j'}}(-1)^{(k_{[d]\setminus\cI})_{j}}
	\inprod{j'}{([d]\setminus\cI)_{j}}
	=
	\sqrt{\ell/d}.
\ee
\hfill$\square$

From Lemma~\ref{lemma:truepos} we see that the parameter $G$ as introduced in Section~\ref{sec:merit}
is given by
\be
	G=\qa.
\ee

%------------------------------------------------------------
\subsection{Forgery probability per symbol}

We look at the following attack scenario.
The attacker observes a valid signature of symbol~$s$.
He owns all $T$ existing public keys.
His aim is to create a forged signature for a symbol $t\in\cS$, with $t\neq s$.
We define
\be
	J\isdef \pr\Big[{\rm forged\;qudit\;gives\;projection\; 0}\Big]
\ee
\be
	\qf\isdef T\frac{\log d}d.
\ee

\begin{lemma}
\label{lemma:EWform}
Consider the forgery of one symbol.
Let the random variable $A\in\bits^{\ell-d}$ be the part of $k$ unavailable to the attacker.
Let the random variable $W\in\{0,\ldots,\ell-d\}$ be the Hamming distance between $A$ and 
the attacker's guess for~$A$.
The success probability for forging one symbol can be expressed as
\be
	1-J = (1-\qa)\EE_W (1-2\frac W\ell)^2.
\label{EWform}
\ee
\end{lemma}
\underline{\it Proof:} 
Let $K$ be the forged signature, with Hamming distance $w$ w.r.t. the correct key $k$. 
We have
\bea
	\inprod{\mu(k)}{\qj(\cI(t),K)} &=&
	\frac1{\sqrt{\ell d}}\sum_{j=0}^{\ell-1}\sum_{j'=0}^{d-1} (-1)^{k_{j'}}(-1)^{K_{\cI(t)_j}}\inprod{j'}{j}
	\nn\\ &=&
	\frac{\mbox{\#correct}-\mbox{\#wrong}}{\sqrt{\ell d}}
	=\frac{\ell-2W}{\sqrt{\ell d}}.
\eea
Squaring and taking the expectation over $W$ yields (\ref{EWform}).
\hfill$\square$

\begin{conjecture}
\label{conj:fdecreasing}
Let $f(x,r)$ be defined as
\be
	f(x,r)\isdef \frac1r\cdot \frac{\sum_{w=0}^r w{x\choose w}}{\sum_{w=0}^r {x\choose w}},
	\quad\quad r\leq \frac x2.
\ee
The function $f(x,r)$ is decreasing in~$r$.
\end{conjecture}

\underline{\it Corroboration:}
We verified this numerically for samples of $x$ up to $x=2^{28}$.
\hfill$\square$

Remark: The range of $x$ for which we tested the validity of Conjecture~\ref{conj:fdecreasing}
entirely covers the numerical results presented in Section~\ref{sec:num}.

\begin{proposition}
\label{prop:p1}
For a non-matching position the accept probability can be bounded as 
\be
	1-J \leq p_1,
\ee
where
\be
	p_1\isdef(1-\qa)\Big[ \Big(1-\frac\qa{1-\qa}2h^{\rm inv}(1-\frac\qf\qa)\Big)^2+\sqrt{\frac83}\cdot \frac{\sqrt{d-\ell}}{\ell}\Big].
\label{forgeryresult}
\ee
\end{proposition}
\underline{\it Proof:}
For any distribution $P$ of $A$ (from the attacker's point of view) the probability of succesful forgery is maximised by
outputting the most likely value of $A$, i.e. $a_0={\rm argmax}_a \pr_{A\sim P}[A=a]$. 
Then $W=|A\oplus a_0|$.
We introduce shorthand notation 
$p_a=\pr[A=a]$ and 
$\pmax=\max_a\pr_{A\sim P}[A=a]$.
From the space of distributions $P$ satisfying the constraint $\Hmin(A)=-\log\pmax$ (with fixed $\pmax$)
we will determine which $P$ maximizes (\ref{EWform}).
For $a\neq a_0$ we parametrise $p_a=\pmax \sin^2 \qy_a\in[0,\pmax]$.
The Lagrangian for the optimisation is
\be
	\cL = \sum_{a:a\neq a_0} \pmax \sin^2\qy_a (1-2\frac{|a\oplus a_0|}\ell)^2
	+\ql[1-\pmax- \sum_{a: a\neq a_0} \pmax \sin^2\qy_a],
\ee
where $\ql$ is a constraint multiplier.
Setting the derivatives w.r.t. $\qy_a$ to zero yields
\be
	0=[(1-2\frac{|a\oplus a_0|}\ell)^2-\ql]\sin 2\qy_a.
\label{derivzero}
\ee
Unless $\ql$ has a special value, (\ref{derivzero}) implies $\qy_a=0 \vee \qy_a=\fr\pi2$ for all $a$,
i.e. $p_a=0 \vee p_a=\pmax$.

Eq.(\ref{EWform}) is maximal when $P$ has the following form:
strings $a$ with low values of $|a\oplus a_0|$ have probability $\pmax$,
whereas strings $a$ with high values of $|a\oplus a_0|$ have probability~$0$.
The nonzero probability is concentrated within a `radius' $r$ around $a_0$,
with $r\leq (d-\ell)/2$.
\be
	\frac1\pmax=\sum_{w=0}^r{d-\ell \choose w}.
\ee
Lemma~\ref{lemma:radius} together with $d-\ell=\qa d$ and $\log\pmax^{-1}=d(\qa-\qf)$ yields
\be
	r \geq (d-\ell)h^{\rm inv}(1-\frac\qf \qa).
\label{rlowerbound}
\ee
From Lemma~\ref{lemma:EWform} we have $1-J=(1-\qa)[1-\frac4\ell\EE_W W+\frac4{\ell^2}\EE_W W^2]$.
Using $W\leq r$ we can write
$1-J\leq (1-\qa)[(1-2r/\ell)^2 +\frac4\ell\EE_W(r-W)]$.
Next, using (\ref{rlowerbound}) we get 
\be
	1-J\leq (1-\qa)\Big[ \Big(1-\frac{\qa}{1-\qa}2h^{\rm inv}(1-\frac\qf \qa)\Big)^2 +\frac{4r}\ell\EE_W(1-\frac Wr)\Big].
\ee
Finally we have to upper bound $\EE_W(1-W/r)$.
\be
	\frac1r\EE_W W=\frac1r\cdot\frac{\sum_{w=0}^r{d-\ell \choose w} w}{\sum_{w=0}^r{d-\ell \choose w}}
	\isdef  f(d-\ell,r).
\ee
Since $f$ is decreasing in $r$ (Conjecture~\ref{conj:fdecreasing}) 
we have
\be
	\EE_W W \geq r\cdot f(d-\ell,\frac{d-\ell}2).
\ee
Without loss of generality we assume that $d-\ell$ is even and write $d-\ell = 2z$.
\bea
	f(2z,z) &=&
	\frac1z 
	\frac{\sum_{w=0}^z{2z \choose w} w}{\sum_{w=0}^z{2z \choose w}}
	\nn\\ &=&
	2 \frac{\sum_{w=0}^{z-1}{2z-1 \choose w}}{\sum_{w=0}^z{2z \choose w}} 
	\nn\\ &=&
	2\frac{\fr12\cdot 2^{2z-1}}{\fr12[2^{2z} +{2z\choose z}]}=
	\frac1{1+{2z\choose z}/2^{2z}}
	\nn\\ & \geq&
	\frac1{1+\frac1{\sqrt{3z+1}}}.
\eea
Thus we have
\be
	\frac4\ell \EE_W(r-W) \leq \frac{4r}{\ell}\frac{[1+\fr32(d-\ell)]^{-1/2}}{1+[1+\fr32(d-\ell)]^{-1/2}}
	< \frac{4r}{\ell} [\fr32(d-\ell)]^{-1/2}.
\ee
With $r\leq (d-\ell)/2$ this gives
\be
	\frac4\ell \EE_W(r-W) < \frac{2\sqrt2}{\sqrt3}\cdot\frac{\sqrt{d-\ell}}\ell.
\ee
\hfill$\square$

Note that Proposition~\ref{prop:p1} invokes Conjecture~\ref{conj:fdecreasing}.

\begin{corollary}
\label{corol:3alpha}
The expression $p_1$ defined in (\ref{forgeryresult}) can be lower bounded as
\be
	p_1> 1-3\qa.
\ee
\end{corollary}

\underline{\it Proof}.
First we write out the square in (\ref{forgeryresult}) and neglect most of the positive terms, yielding
$p_1> 1-\qa-2\qa\cdot 2h^{\rm inv}(1-\frac\qf\qa)$. 
Then we use $h^{\rm inv}(\cdot)\leq \fr12$.
\hfill$\square$

%----------------------------------------
\subsection{Setting the parameters; asymptotics}

Let $d=2^n$, which means that each qudit can be thought of as $n$ qubits.
The $\qf$ and $1/d$ are exponentially small in $n$.
For large $d$ one can set $\qa\ll 1$, $\qf/\qa\ll 1$.
Then $J\approx 1-p_1$ (\ref{forgeryresult}) equals $\qa+2\qa\cdot2h^{\rm inv}(1-\frac\qf\qa)$ minus higher order terms.
In contrast, a genuine signature has per-qudit error probability $G=\qa$.

The scheme must allow the verifier to distinguish between error rate $\qa$ and $J$.

\begin{proposition}
\label{prop:settings}
Consider the scheme proposed in Section~\ref{sec:steps}, with 
parameters $\qy,N,z_{\rm acc},z_{\rm rej}$ set as follows
as a function of $\qa,d,T$,
\bea
	\qy &=& \frac\qa2(1+\nu)\quad\mbox{with }\nu\mbox{ constant}>\qa  
\label{settheta}
	\\
	N &=& \frac1{\qa^3}\left(\frac{\sqrt{3\ln \qe_{\rm c}^{-1}}
	+\sqrt{1+4\qy}\sqrt{2\ln \qe_{\rm f}^{-1}}}
	{\frac{1-p_1-\qa}\qa}\right)^2
\label{setN}
	\\
	z_{\rm acc}&=& N\qa+\sqrt{N\qa}\sqrt{3\ln \qe_{\rm c}^{-1}}
\label{setzacc}
	\\
	z_{\rm rej} &=& (1-2\qy)N\qa+2\qy N(1-p_1)
	\nn\\ && 
	-\sqrt{(1-2\qy)N\qa+2\qy N(1-p_1)}\sqrt{2\ln \qe_{\rm f}^{-1}}
\label{setzrej}
\eea
with $p_1$ as defined in (\ref{forgeryresult}).
Given these settings the scheme is correct with error $\qe_{\rm c}$ as specified in Def.\,\ref{def:correctness}
and $\qe_{\rm f}$-secure against forgery as specified in Def.\,\ref{def:forgery}.
\end{proposition}

\underline{\it Proof.}
Peggy's signature has error probability $\qa$ in each individual qudit.
The expected tally of errors in the codeword in $N\qa$.
By Lemma~\ref{lemma:Chernoff} and the setting of $z_{\rm acc}$ (\ref{setzacc}) 
the probability that the tally exceeds $z_{\rm acc}$ is upper bounded by
$\qe_{\rm c}$.

The `minimal' forgery consists of modifying one symbol in the message $x$.
Let $\tilde x\in\cS^K$ be the modified message and let $\tilde c\in\cS^N$ be the corresponding codeword.
The Hamming distance between $\tilde c$ and $c$ is $2\qy N$.
There are $N(1-2\qy)$ symbols that the attacker does not have to modify; 
in these positions the error rate is~$\qa$.
In the $2\qy N$ positions that the attacker must modify he introduces an error rate~$J$.
Hence the expected overall error tally for the forgery is $E=(1-2\qy) N\cdot\qa+2\qy NJ$
$=N\qa+N\cdot2\qy(J-\qa)$.
Eq.~(\ref{setzrej}) has the form $E-\sqrt E\sqrt{2\ln \qe_{\rm f}^{-1}}$;
substitution into Lemma~\ref{lemma:Chernoff} yields the correct bound on the forgery probability.

We have to enforce $z_{\rm rej}<2\qy N$. 
We do this by ensuring that $N\qa+N\cdot2\qy(1-p_1-\qa) <2\qy N$.
This condition can be written as
$2\qy(1-[1-p_1-\qa])> \qa$, which is satisfied because of (\ref{settheta})
and $p_1<1$.

Finally we have to enforce $z_{\rm rej}>z_{\rm acc}$.
Setting $N$ as in (\ref{setN}) ensures that this condition is satisfied, as can be verified by
a straightforward but tedious computation.
\hfill$\square$

\vskip2mm

Proposition~\ref{prop:settings} depends on Conjecture~\ref{conj:fdecreasing}.

Note in (\ref{setN}) that asymptotically $N$ is of order $\qa^{-3}$.
Setting $\qa$ to be small has advantages, but these advantages can be
exploited only if the signed message is sufficiently long.

%-------------------------------------------------
\subsection{Non-repudiation}
\label{sec:nonrepud}

\begin{lemma}
\label{lemma:schemerepud}
With the parameter settings given in Proposition~\ref{prop:settings},
our scheme is $\qe$-secure against non-repudiation as defined in
Def.\,\ref{def:nonrepudiation},
with $\qe=T\cdot \max(\qe_{\rm f}^{\nu^2},\qe_{\rm c}^{\nu^2})\cdot[1+\cO(\qa)]$.
\end{lemma}
\underline{\it Proof sketch:}
Malicious Peggy can prepare any state, but all $T$ verifiers receive the same state.
Peggy's best chance of causing Rejects as well as 1-Accepts at the different verifiers
is to (i) fix the message $x$ before distributing the public key and then 
(ii) as the `public key' in each of the $N$ positions prepare a state that is tuned to cause 
error probability $J$, with $J=(z_{\rm acc}+z_{\rm rej})/(2N)$.
We use Lemma~\ref{lemma:Chernoff} with $\mu\to (z_{\rm acc}+z_{\rm rej})/2$, 
$\qd\to (z_{\rm rej}-z_{\rm acc})/(z_{\rm acc}+z_{\rm rej})$ to obtain
\be
	\pr[Z\leq z_{\rm acc}] \leq e^{-\fr14\frac{(z_{\rm rej}-z_{\rm acc})^2}{z_{\rm acc}+z_{\rm rej}}},
	\quad
	\pr[Z\geq z_{\rm rej}] \leq e^{-\fr14\frac{(z_{\rm rej}-z_{\rm acc})^2}{z_{\rm acc}+z_{\rm rej}}}
	\cdot[1+\cO(\qa)].
\label{Zbound}
\ee
From Proposition~\ref{prop:settings} we get
\bea
	z_{\rm rej}-z_{\rm acc} & \!=\! & \frac{2\qy/\qa-1}{1-p_1-\qa} \!\!
	\left(\sqrt{3\ln \qe_{\rm c}^{-1}}+\sqrt{2\ln \qe_{\rm f}^{-1}}\right)^2
	\!\! [1+\cO(\qa)]
	\quad\quad
	\\
	z_{\rm rej}+z_{\rm acc} & = & \frac2\qa
	\left(\frac{\sqrt{3\ln \qe_{\rm c}^{-1}}+\sqrt{2\ln \qe_{\rm f}^{-1}}}{1-p_1-\qa}\right)^2
	[1+\cO(\qa)].
\eea
Substitution into (\ref{Zbound}) yields the same expression for both bounds,
\be
	\exp[-\frac{\nu^2}8\left(\sqrt{3\ln \qe_{\rm c}^{-1}}+\sqrt{2\ln \qe_{\rm f}^{-1}}\right)^2]
	\leq
	\max( \qe_{\rm f}^{\nu^2}, \qe_{\rm c}^{\nu^2}).
\ee
Finally we use (\ref{Prepudineq2}) in Corollary~\ref{corol:repud}.
\hfill$\square$

Note that the bound in Lemma~\ref{lemma:schemerepud} is not tight.
We present the lemma mainly to show that the repudiation probability is under control.
A more complete treatment of non-repudiation, instead of merely a heuristic,
is left for future work.

%------------------------------------------------------
\subsection{How many qubits of public key are spent per signed message bit}
\label{sec:qubitsperbit}

We show that the size of our public key, taken per signed message bit,
can be significantly smaller than in GC01.
Expressed in qubits, our public key has size $N\log d$.
The length of the message is $K$ nonbinary symbols, which is equivalent to
$K\log S$ bits.
Asymptotically $K\to N[1-h(\qy)]$, where $\qy\approx \qa/2$ if the parameters are set
according to Proposition~\ref{prop:settings}.
Thus we can write
\be
	\frac{\mbox{\#qubits}}{\#\rm bits}= \frac{N\log d}{K\log S}
	\approx\frac{\log d}{\log S}\cdot\frac1{1-h(\qy)}.
\label{expenditure1}
\ee
Furthermore, we have the requirement $d-\ell\geq T\log d$, since the number of unknown bits
in $k$ must not be smaller than what can be learned from $T$ copies of a $d$-dimensional qudit.
This requirement can be rewritten as $d\geq ST\log d$.
Substitution into (\ref{expenditure1}) gives
\be
	\frac{\mbox{\#qubits}}{\#\rm bits}\geq (1+\frac{\log T+\log\log d}{\log S})\frac NK
	\approx
	(1+\frac{\log T+\log\log d}{\log S})\frac1{1-h(\qy)}.
\ee
Compared to the GC01 expenditure (\ref{GCsmallgamma})
our scheme is more efficient by a factor of roughly $\log S$.
In theory it is possible to set $S\gg T$ to obtain a public key
whose size is just slightly more than one qubit per signed message bit.

%----------------------------------------------------------
\subsection{Numerics}
\label{sec:num}

We numerically compare our scheme against GC01-with-codewords.
We assume an efficient form of GC01 that re-uses unspent quantum states,
i.e.~if $\ket{P^0_j}$ is measured then 
$\ket{P^1_j}$ gets relabeled in some way and re-used later.

We fix the number of verifiers $T$ 
and the error parameter $\qe_{\rm f}$. 
We focus on three performance indicators:
(i) how many qubits are spent per signed message bit,
(ii)~the gap $J-G$, and
(iii)~the codeword length~$N$.

\vskip2mm

\begin{table}
\centering
\caption{\it Comparison between GC01 and our scheme.}
{\small
\begin{tabular}{|l|l|l|l|l|l|}
\hline
& \!\!tunable\!\!\! & \!qubits/bit\! & $1-J$ & $N$ & gap 
\\ \hline
\!\!GC\!\! & $d,\qg,\qb$ & $\frac{\log d}{1-h(\qb)}$ &
$\frac{1-(1-4\qg)^2}{2^{d[1-h(\qg)-\qf]}} +(1-4\qg)^2 $
&
$\frac{\ln (1-Q_{\rm R}^{\rm forgery})^{-1}}{\qb J^{\rm GC}}$
&
$J^{\rm GC}$
\\ \hline
us & $d,\qa,\qy$ & 
$\frac{\log d / \log \fr1\qa}{1-h(\qy)}$\!\!
& 
$p_1 = (1-\qa)\cdot$
& 
$\cO(\frac1{\qa(J-G)^2})$
&
\!\!$1-p_1-\qa$\!\!
\\ 
&&&
$\Big[ \Big(1-\frac\qa{1-\qa}2h^{\rm inv}(1-\frac\qf\qa)\Big)^2+\sqrt{\frac{8\qa}{3d}}\cdot \frac{1}{1-\qa} \Big]$
&
(\ref{setN})
&
\\
\hline
\end{tabular}
}
\end{table}

\vskip2mm

\begin{figure}[!h]
\begin{center}
\includegraphics[width=100mm]{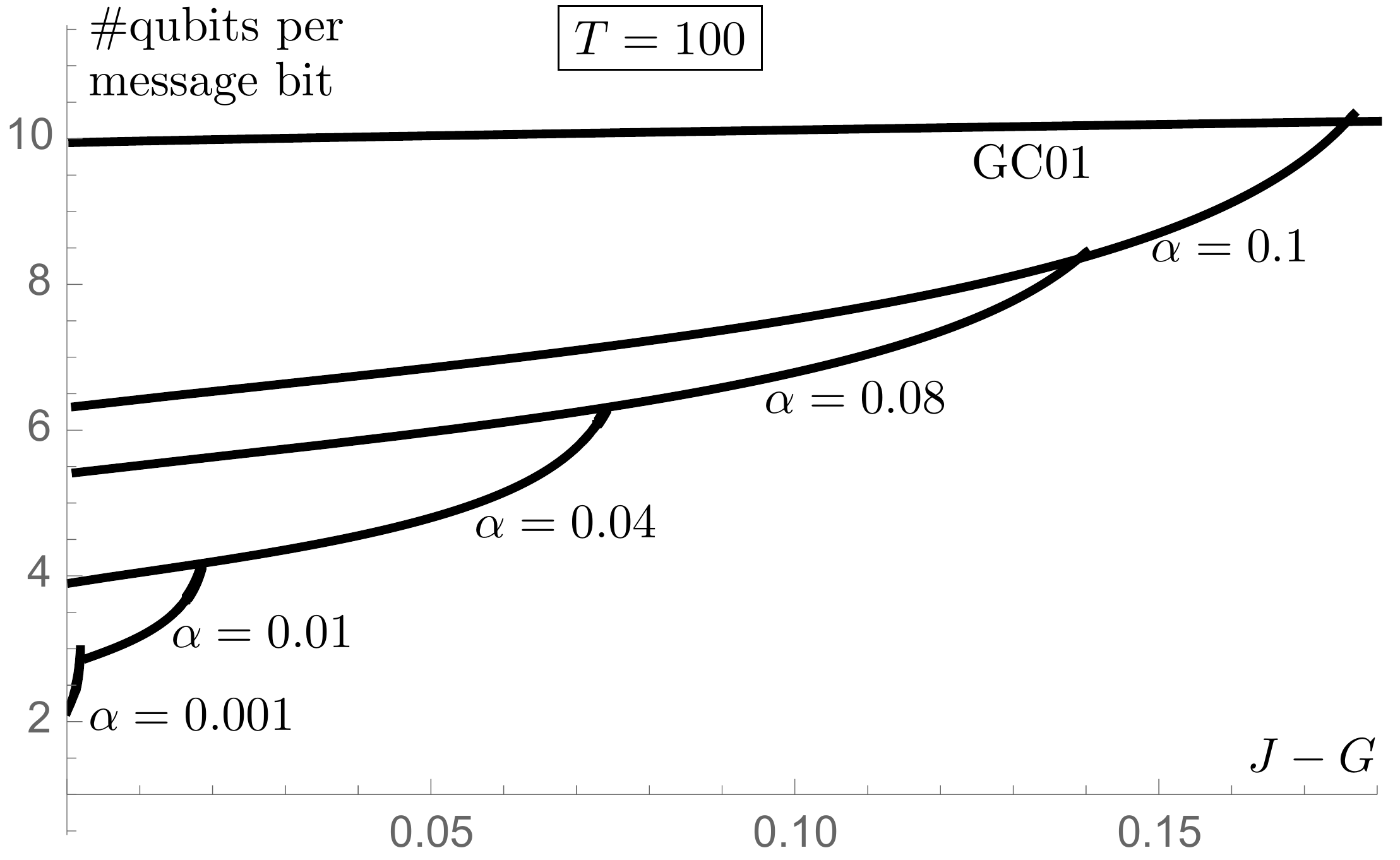}
\end{center}
\caption{\it
Number of spent qubits per signed message bit versus the `gap' $J-G$, at $T=100$,
plotted at various values of~$\qa$.
The almost horizontal curve is the Gottesman-Chuang scheme.
In each curve the parameter $d$ is varied.
The ranges of $d$ are roughly as follows.
At $\qa=0.001$: $d\in(3\cdot 10^6, 8\cdot10^8)$;
at $\qa=0.01$: $d\in(3\cdot10^5,8\cdot10^7)$;
at $\qa=0.04$: $d\in(4\cdot10^4,3\cdot10^7)$;
at $\qa=0.08$: $d\in(2\cdot10^4,5\cdot10^6)$;
at $\qa=0.1$: $d\in(2\cdot10^4,7\cdot10^6)$.
}
\label{fig:T100}
\end{figure}

\begin{figure}[!h]
\begin{center}
\includegraphics[width=100mm]{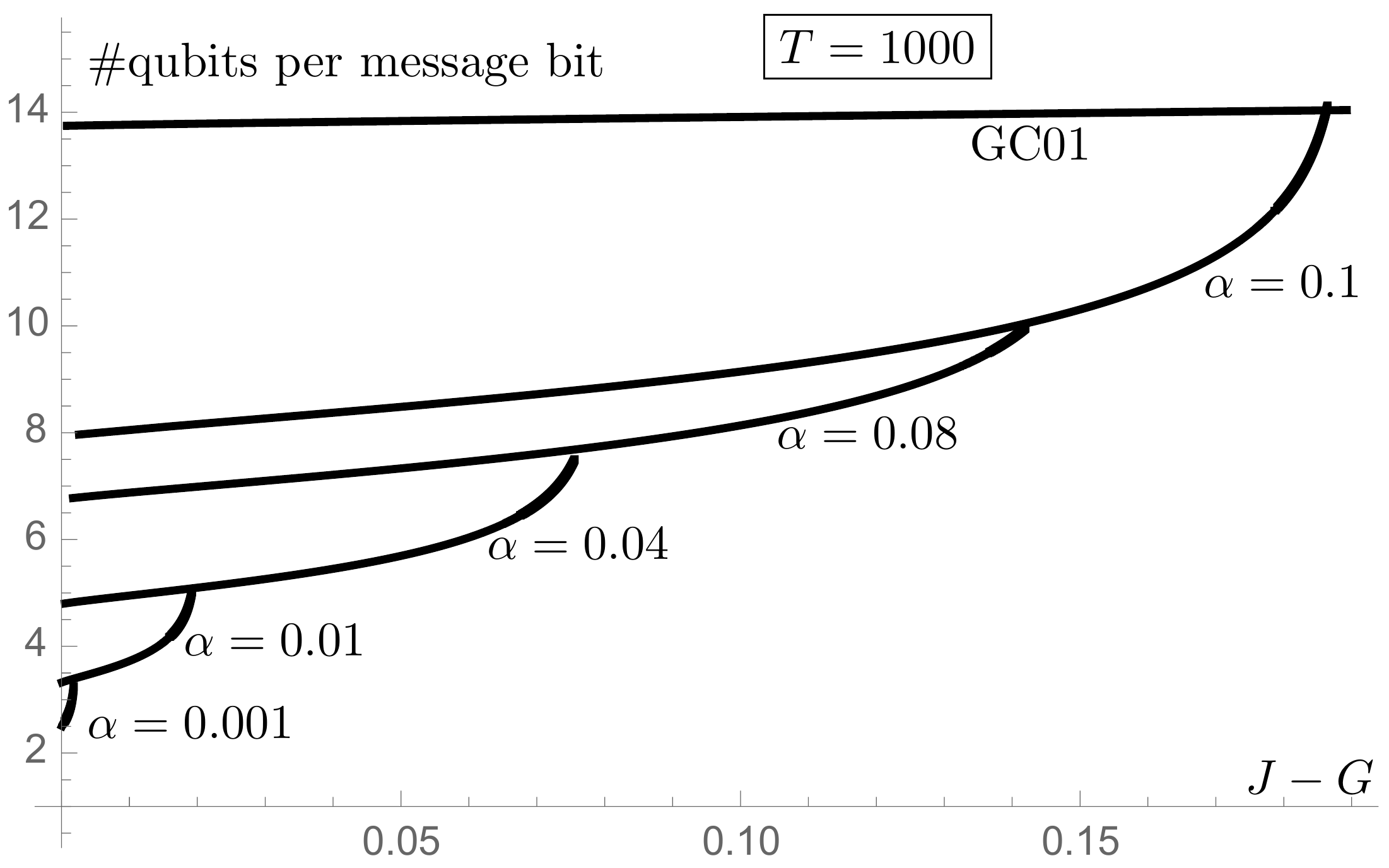}
\end{center}
\caption{\it
Number of spent qubits per signed message bit versus the `gap' $J-G$, at $T=1000$,
plotted at various values of~$\qa$.
The almost horizontal curve is the Gottesman-Chuang scheme.
In each curve the parameter $d$ is varied.
The ranges of $d$ are roughly as follows.
At $\qa=0.001$: $d\in(3\cdot 10^7, 5\cdot10^9)$;
at $\qa=0.01$: $d\in(2\cdot10^6,4\cdot10^9)$;
at $\qa=0.04$: $d\in(5\cdot10^5,8\cdot10^8)$;
at $\qa=0.08$: $d\in(2\cdot10^5,8\cdot10^7)$;
at $\qa=0.1$: $d\in(2\cdot10^5,2\cdot10^9)$.
}
\label{fig:T1000}
\end{figure}

\begin{figure}[!h]
\begin{center}
\includegraphics[width=100mm]{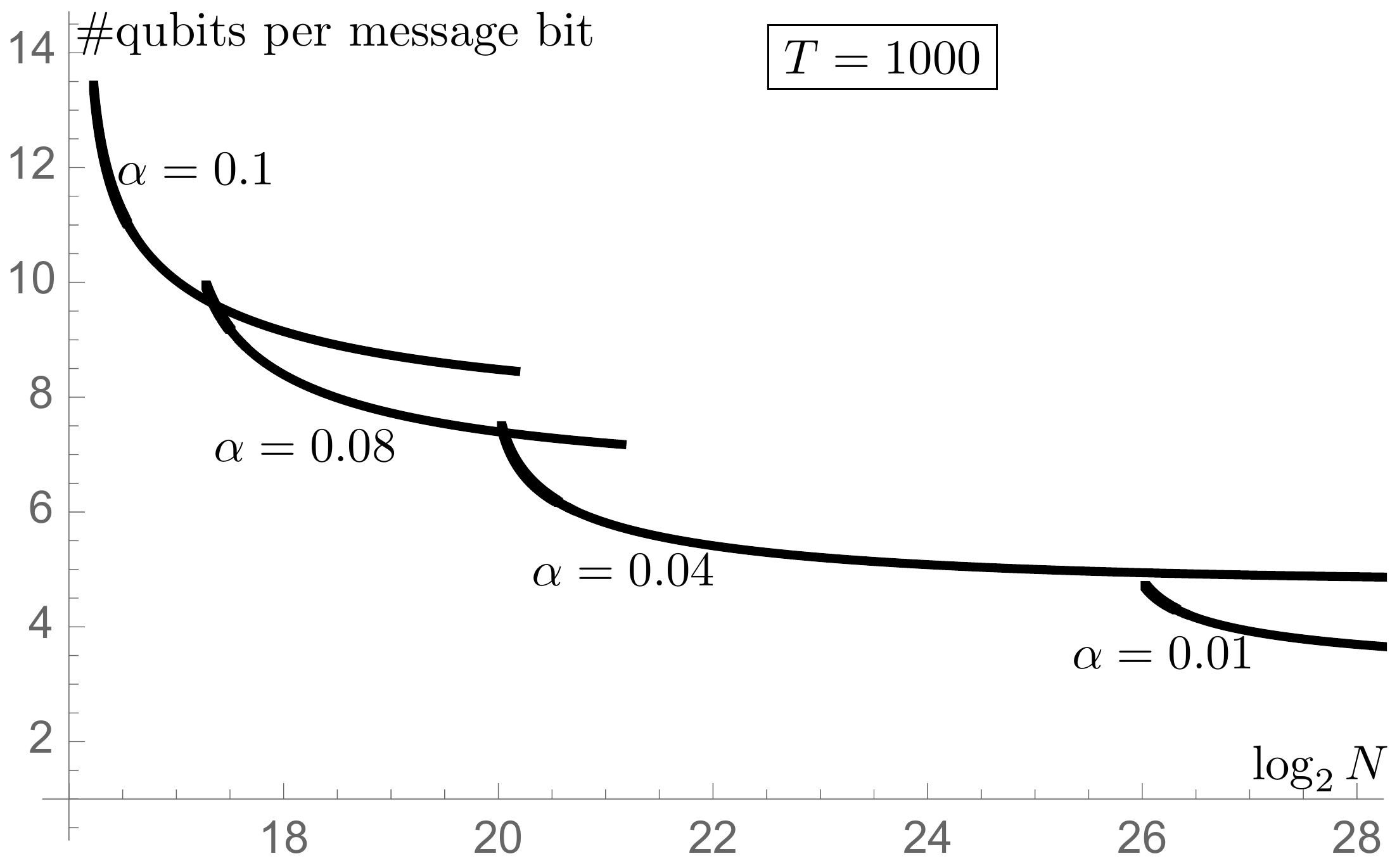}
\end{center}
\caption{\it
Number of spent qubits per signed message bit versus the codeword length~$N$ (in bits),
for $T=1000$ and various values of~$\qa$.
$\qe_{\rm f}=10^{-12}$; 
$\qe_{\rm c}=10^{-9}$.
In each curve the parameter $d$ is varied.
}
\label{fig:codelength}
\end{figure}

Figs.\,\ref{fig:T100} and \ref{fig:T1000}
show plots of the qubit expenditure versus the gap $J-G$ for several combinations
of $T$ and~$\qa$, with parameter settings as in Proposition~\ref{prop:settings}. 
At small values of $\qa$ (large alphabet size $S$) the public keys are significantly smaller
than for GC01.
Fig.\,\ref{fig:codelength} has the code length $N$ on the horizontal axis.
We see that (for the chosen parameter settings) the gain over GC01 sets in when
the message length is of the order of magnitude of 8 kilobytes.

%\clearpage

%======================================================
\section{Discussion}
\label{sec:summary}

It may be possible to improve on the parameter settings given in
Proposition~\ref{prop:settings}, and on some of the bounds that we have derived, 
e.g.\,by using tighter concentration inequalities.

Our treatment of non-repudiation is less general than that of GC01, who allow
Peggy to distribute more general states that pass swap tests.
A full treatment would entail determining how large the parameter $T$ (the number of copies) needs to be,
as a function of $\qa$, in order
to ensure that the swap tests sufficiently reduce Peggy's probability of distributing different
public keys.
Note that Proposition~\ref{prop:settings} allows a lot of freedom for choosing~$T$;
hence we expect that the required $T$ can be accommodated.
This analysis is left for future work.

It is rather embarrassing that we do not have an actual proof for Conjecture~\ref{conj:fdecreasing}
but only numerical evidence.
Fortunately, it is rather straightforward (though time consuming) to verify numerically that
the conjecture holds for very large values of~$x$.
The graphs plotted in Section~\ref{sec:num} do not exceed the verified range of~$x$.
Hence it is clear that there is a wide parameter regime in which our scheme is advantageous

If Conjecture~\ref{conj:fdecreasing} holds,
our scheme asymptotically achieves a public key size of one qubit per message bit.
The question naturally arises whether it is possible to go below that value,
and if a theoretical lower bound exists.
Consider a set of $N=2^A$ ordinary GC01 public keys labeled ${0,\ldots,N-1}$, and let the opening of 
the key with label $x$ represent a signature of the $A$-bit binary message~$x$.
The expended key gets replaced by a new one, while all the other public keys remain in use.
Such a scheme spends $(\log d)/A$ qubits per message bit, which can definitely be smaller than~1;
however, it requires more complicated synchronisation between the prover and the verifiers 
than the scheme presented in this paper.

%=======================================================================

%\clearpage

\section*{Acknowledgements}
%\vskip-2mm
We thank Ronald de Wolf, Andreas H{\"u}lsing and Aart Blokhuis for useful discussions.

\newpage

%============================================================
\bibliographystyle{unsrt}
\bibliography{qsig_arx2}

%========================================================================
\end{document}